\documentclass{JHEP}

\usepackage{amsfonts,amsbsy}

\newcommand{\be}{\begin{equation}}
\newcommand{\ee}{\end{equation}}

\newcommand{\UN}{U(N)}

\newcommand{\ZN}{\mathbf{Z}_N}

\newcommand{\Dsl}{\not\!\!D}

\newcommand{\bea}{\begin{eqnarray}}
\newcommand{\eea}{\end{eqnarray}}

\newread\epsffilein    
\newif\ifepsffileok    
\newif\ifepsfbbfound   
\newif\ifepsfverbose   
\newdimen\epsfxsize    
\newdimen\epsfysize    
\newdimen\epsftsize    
\newdimen\epsfrsize    
\newdimen\epsftmp      
\newdimen\pspoints     
\def\epsfbox#1{\global\def\epsfllx{72}\global\def\epsflly{72}%
   \global\def\epsfurx{540}\global\def\epsfury{720}%
   \def\lbracket{[}\def\testit{#1}\ifx\testit\lbracket
   \let\next=\epsfgetlitbb\else\let\next=\epsfnormal\fi\next{#1}}%
\def\epsfgetlitbb#1#2 #3 #4 #5]#6{\epsfgrab #2 #3 #4 #5 .\\%
   \epsfsetgraph{#6}}%
\def\epsfnormal#1{\epsfgetbb{#1}\epsfsetgraph{#1}}%
\def\epsfgetbb#1{%
\openin\epsffilein=#1
\ifeof\epsffilein\errmessage{I couldn't open #1, will ignore it}\else
   {\epsffileoktrue \chardef\other=12
    \def\do##1{\catcode`##1=\other}\dospecials \catcode`\ =10
    \loop
       \read\epsffilein to \epsffileline
       \ifeof\epsffilein\epsffileokfalse\else
          \expandafter\epsfaux\epsffileline:. \\%
       \fi
   \ifepsffileok\repeat
   \ifepsfbbfound\else
    \ifepsfverbose\message{No bounding box comment in #1; using defaults}\fi\fi
   }\closein\epsffilein\fi}%
\def\epsfclipstring{}
\def\epsfsetgraph#1{%
   \epsfrsize=\epsfury\pspoints
   \advance\epsfrsize by-\epsflly\pspoints
   \epsftsize=\epsfurx\pspoints
   \advance\epsftsize by-\epsfllx\pspoints
   \epsfxsize\epsfsize\epsftsize\epsfrsize
   \ifnum\epsfxsize=0 \ifnum\epsfysize=0
      \epsfxsize=\epsftsize \epsfysize=\epsfrsize
      \epsfrsize=0pt
     \else\epsftmp=\epsftsize \divide\epsftmp\epsfrsize
       \epsfxsize=\epsfysize \multiply\epsfxsize\epsftmp
       \multiply\epsftmp\epsfrsize \advance\epsftsize-\epsftmp
       \epsftmp=\epsfysize
       \loop \advance\epsftsize\epsftsize \divide\epsftmp 2
       \ifnum\epsftmp>0
          \ifnum\epsftsize<\epsfrsize\else
             \advance\epsftsize-\epsfrsize \advance\epsfxsize\epsftmp \fi
       \repeat
       \epsfrsize=0pt
     \fi
   \else \ifnum\epsfysize=0
     \epsftmp=\epsfrsize \divide\epsftmp\epsftsize
     \epsfysize=\epsfxsize \multiply\epsfysize\epsftmp   
     \multiply\epsftmp\epsftsize \advance\epsfrsize-\epsftmp
     \epsftmp=\epsfxsize
     \loop \advance\epsfrsize\epsfrsize \divide\epsftmp 2
     \ifnum\epsftmp>0
        \ifnum\epsfrsize<\epsftsize\else
           \advance\epsfrsize-\epsftsize \advance\epsfysize\epsftmp \fi
     \repeat
     \epsfrsize=0pt
    \else
     \epsfrsize=\epsfysize
    \fi
   \fi
   \ifepsfverbose\message{#1: width=\the\epsfxsize, height=\the\epsfysize}\fi
   \epsftmp=10\epsfxsize \divide\epsftmp\pspoints
   \vbox to\epsfysize{\vfil\hbox to\epsfxsize{%
      \ifnum\epsfrsize=0\relax
        \includegraphics{#1}%
      \else
        \epsfrsize=10\epsfysize \divide\epsfrsize\pspoints
        \includegraphics{#1}%
      \fi
      \hfil}}%
\global\epsfxsize=0pt\global\epsfysize=0pt}%
{\catcode`\%=12 \global\let\epsfpercent=
\long\def\epsfaux#1#2:#3\\{\ifx#1\epsfpercent
   \def\testit{#2}\ifx\testit\epsfbblit
      \epsfgrab #3 . . . \\%
      \epsffileokfalse
      \global\epsfbbfoundtrue
   \fi\else\ifx#1\par\else\epsffileokfalse\fi\fi}%
\def\epsfempty{}%
\def\epsfgrab #1 #2 #3 #4 #5\\{%
\global\def\epsfllx{#1}\ifx\epsfllx\epsfempty
      \epsfgrab #2 #3 #4 #5 .\\\else
   \global\def\epsflly{#2}%
   \global\def\epsfurx{#3}\global\def\epsfury{#4}\fi}%
\def\epsfsize#1#2{\epsfxsize}
\let\epsffile=\epsfbox

\preprint{FTUAM-98-13; IFT-UAM/CSIC-98-10} 

\title{\bf Nahm's Transformation on the Lattice}

\author{ A. Gonz\'alez-Arroyo\thanks{also at: \  Instituto de F\'{\i}sica
Te\'orica C-XVI, Universidad Aut\'onoma de Madrid, Cantoblanco, Madrid~28049,
 SPAIN.} \ and C. Pena \\ Departamento de F\'{\i}sica Te\'orica C-XI\\
Universidad   Aut\'onoma de Madrid\\Cantoblanco, Madrid 28049, SPAIN.}

\abstract{ By studying zero modes of the Dirac operator on the lattice, 
 we explicitly construct the Nahm transform of some topologically 
 non-trivial gauge field configurations on the torus.}


\begin{document}


\section{Introduction}
Classical Yang-Mills fields have been the subject of intense study
in the literature.  Besides being interesting objects in their own right,
much of the motivation  arises from the relevance of non-abelian
gauge theories for our understanding of the interactions among elementary
particles. Although, in that case, one is concerned with  quantum fields,
 particular classical configurations have been argued  to play a fundamental
role in explaining some phenomena. In this context, the pioneer
 work of Polyakov~\cite{Polyakov1} showed how pseudoparticle solutions
 become relevant in the semiclassical approximation to the path integral.
 These solutions are then  argued to be responsible of some
of the most intriguing effects of Quantum Field Theory. This is the case
for the Confinement property of the 3-dimensional compact abelian model,
explained  originally by Polyakov. In 4 dimensions `t~Hooft~\cite{thooft1}
showed that instantons~\cite{BPST}
provide an explanation  for the so-called $U(1)$-problem. This triggered a
joint effort of physicists and mathematicians in the 70's, which culminated
with the construction of all self-dual solutions to the euclidean Yang-Mills
equations compactified on a sphere~\cite{ADHM}. On physical grounds, the
compactification on a sphere is a condition equivalent to the requirement of
finite classical action. In other words, one is looking for solutions which
exist isolated, i.e. surrounded by the perturbative vacuum ($F_{\mu \nu}=0$).
However, in Quantum Field Theory neither the requirement of finite action
nor the one of isolated solutions is compulsory. For example, it was
suggested by the work of Saviddy~\cite{saviddy} and the Copenhagen
group~\cite{copen} that the QCD  vacuum was permeated by some
structures carrying chromo-magnetic flux. These are neither isolated nor carry
finite action. Henceforth, it could  also be relevant and interesting to study
classical solutions with different boundary conditions. 

In this respect a
good deal of interest has focused on the study of Yang-Mills fields on the
4-dimensional torus (for a review see Ref.~\cite{torusrev}).  The torus adds
topological features which have an appealing physical
interpretation~\cite{twist}. Furthermore, solutions on the torus can be
considered solutions on ${\bf R}^4$ which are periodic, with a total action
which diverges (with finite action per unit cell). Unfortunately, despite the
effort, only very special self-dual solutions on the torus, having constant
field strength, are known
analytically~\cite{stsds,vanbaal6}.
The lattice formulation of Yang-Mills theory has proven a  precise and
efficient method for obtaining these solutions numerically~\cite{gpg-as,gpg-a}.
A basic building block of many of these configurations is a certain lump
carrying fractional topological charge $Q=\frac{1}{N}$. These lumps cannot
exist isolated ---their size is determined by the distance to their
neighbours. The moduli space of a self-dual solution on the torus, given by
the index theorem, has four parameters per lump, which can be associated to
the coordinates of its center. This  suggests that one can obtain self-dual
solutions on ${\bf R}^4$ by deforming these configurations away from the
periodic arrangement (at no cost of action). A picture of the Yang-Mills
quantum vacuum as a liquid of these fractional topological charge lumps,
as suggested by our group~\cite{investigating}, could account at the same time
for the observed string tension  and topological susceptibility.

On the analytical side, an interesting tool is provided by  the Nahm
transformation~\cite{nahm}. This transformation maps a $\UN$ self-dual gauge
field on the torus  with topological charge $Q$ into a $U(Q)$ self-dual gauge
field on the dual torus with topological charge $N$. The use of this
transformation, which can be considered a particular case of a duality
transformation, allows one to prove that, in the absence of twist,
there are no $Q=1$ instanton solutions
on the torus (see for example~\cite{pbpvb}). Its use
has proved useful as well for constructing new  instanton solutions in
$S_1\times {\bf
R}^3$ ~\cite{vanbaalnew}. For the torus case, it has been shown that the
constant field strength solutions in $SU(2)$ are mapped into themselves~\cite{vanbaalNT}.
In this paper we will show that it is possible to use lattice techniques to
explicitly construct the Nahm transform of a given self-dual lattice gauge
field configuration. This opens the door to the possibility of systematically
investigating
the properties of  the Nahm transformation for the whole set of
self-dual gauge field configurations.  In the paper we will apply the method
to the case of one of the non-trivial self-dual solution on the torus,
producing accurate results for its transform. As a matter of fact, this
together with the study of the  zero modes for the Dirac
equation in the
adjoint representation,  which are supersymmetric partners of the gauge
field, provides new numerical descriptions of the torus self-dual gauge
fields, with some relative advantages over the conventional discretization.
There is also  hope that, as in the aforementioned $S_1 \times {\bf R}^3$
case,
this study could eventually lead to an analytical 
approach to these classical fields.

The layout of the paper is as follows. In the next section, we present the
details of the Nahm transformation and the main formulas which relate our problem
to the solution of homogeneous and inhomogeneous Dirac equations. In the
following  
section, we explain how one can formulate the latter  on the lattice, and
describe the numerical techniques that we use to solve them. In section 4, we
apply the technique to the actual construction of the Nahm transform for 
some input self-dual configurations. We first test the method by looking 
at its result in a known case, and then apply it to other non-trivial self-dual
configuration. Finally, in section 5 we summarize our results and discuss  
future prospects.

\section{The Nahm transformation}
Let us consider a  4 dimensional torus of size $l_0 \times l_1 \times l_2
\times l_3$, and let $\hat{l}_{\mu}$ represent the vector $(0,\ldots,
l_{\mu},\ldots,0)$, whose only non-zero component is the  $\mu^{th}$ component.
Now consider a self-dual gauge field configuration  $A_{\mu}(x)$ 
defined on this torus. It satisfies:
\be
\label{tbc}
A_{\nu}(x + \hat{l}_{\mu}) = \Omega_{\mu}(x)\, A_{\nu}(x)\, \Omega^+_{\mu}(x) +
\imath\ \Omega_{\mu}(x)\,  \partial_{\nu} \Omega^+_{\mu}(x)\ \ ,
\ee
where $\Omega_{\mu}(x)$ are the twist matrices. For $SU(N)$, these matrices must fulfill the consistency
condition:
\be
\Omega_{\mu}(x+\nu) \Omega_{\nu}(x) = Z_{\mu \nu}\ \Omega_{\nu}(x+\mu) \Omega_{\mu}(x)\ \ .
\ee
When fields transforming in the fundamental representation of the gauge group appear, 
all the constants $Z_{\mu \nu}$ must be equal to 1; otherwise, they are in general
elements of the center of the group $\ZN$, which can be
parametrized as $Z_{\mu \nu}=\exp(2 \pi \imath n_{\mu \nu}/N)$. The twist tensor $n_{\mu \nu}$
is antisymmetric, and its elements are integers defined modulo $N$.

Let $Q$ stand for the
topological charge of the gauge field configuration. As is well-known,
the Atiyah-Singer index theorem implies that the difference between the
number of positive chirality and negative chirality solutions of the Dirac
equation for fermion fields transforming in the fundamental representation of
the gauge group is given by $Q$. Let us assume that for our gauge configuration
there are no  negative chirality
solutions.  Then,  there are exactly $Q$ positive chirality solutions, which we
will label $\Psi^{\alpha}(x)$, with $\alpha=1,\ldots ,Q$. They  satisfy:
\be
\hat{\bar{D}}\Psi^{\alpha}(x)=0\ \ ,
\ee
where $\hat{\bar{D}} \equiv D_{\mu} \bar{\Gamma}_{\mu}$ is the positive chirality Weyl
operator and $D_{\mu} = \partial_{\mu} - \imath A_{\mu}$. In the Weyl basis we have:
\be
\Dsl = \pmatrix{ 0 & \hat{D} \cr  \hat{\bar{D}} & 0 }
\hspace{0.5 cm}
\gamma_5 = \pmatrix{ 1 & 0 \cr  0 & -1 }  \quad ,
\ee
where $\Gamma_{\mu}=(I, -\imath \vec{\sigma})$ and
$\bar{\Gamma}_{\mu}=\Gamma^+_{\mu}$. Furthermore, the solutions satisfy the
following boundary conditions:
\be
\label{bc}
\Psi^{\alpha}(x + \hat{l}_{\mu}) =  \Omega_{\mu}(x) \Psi^{\alpha}(x)\ \ .
\ee

Now  consider the family of gauge fields:
\be
A_{\mu}(x,z)=A_{\mu}(x)+ 2 \pi z_{\mu}\, I ,
\ee
where $z_{\mu}$ are 4 real numbers. For all $z$, the field strength $F_{\mu
\nu}$ is the same, and hence they are all self-dual and have the same
topological charge. Therefore, we obtain a family $\Psi^{\alpha}(x,z)$ 
of positive chirality
solutions of the Dirac equation:
\be
\label{equD}
\hat{\bar{D}}_z \Psi^{\alpha}(x,z) = (\hat{\bar{D}}- 2 \pi \imath \hat{\bar{z}}\,
I) \Psi^{\alpha}(x,z) = 0 \ \ ,
\ee
satisfying the
boundary condition Eq.~\ref{bc} and normalized as follows:
\be
\label{normalization}
\int d^4x\ \left(\Psi^{\alpha}(x,z)\right)^+\,  \Psi^{ \beta}(x,z) = \delta_{\alpha \beta}
\ \ .
\ee
Now notice that  $\exp(- 2 \pi \imath \tilde{z}_{\mu} x_{\mu})\,
\Psi^{\alpha}(x,z+\tilde{z})$ satisfies the same equation than
$\Psi^{\alpha}(x,z)$. However, in general, the boundary conditions are
different, since 
 the right-hand side of Eq.~\ref{bc} gets multiplied by $\exp (-  2 \pi
\imath \tilde{z}_{\mu} l_{\mu})$. This  new factor becomes simply unity if
$\tilde{z}_{\mu}$ is an integer multiple of $1/l_{\mu}$. Hence, defining  the
vector $\hat{\tilde{l}}_{\mu} = (0,\ldots, \frac{1}{l_{\mu}},\ldots,0)$ we can
write:
\be
\label{tbcfors}
\Psi^{\alpha}(x,z+\hat{\tilde{l}}_{\mu}) = \Psi^{\beta}(x,z) (\Omega'^{+}_{\mu}(z))_{\beta \alpha} \exp(2 \pi \imath x_{\mu}/l_{\mu})\ .
\ee
This is so because any solution can be written as a linear combination of the
basis  functions $\Psi^{\beta}(x,z)$. The coefficients
$(\Omega'^{+}_{\mu}(z))_{\beta \alpha}$ cannot in general be chosen equal to
1, if we insist in $\Psi^{\beta}(x,z)$ being continuous in $z$.

Now let us construct the Nahm transform of the gauge field $A_{\mu}(x)$. 
It is given by:
\be
\label{NT}
(\hat{A}_{\mu}(z))_{\alpha \beta} = \imath\, \int d^4x\ \left(\Psi^{
\alpha}(x,z)\right)^+\ \frac{\partial}{\partial z_{\mu}}\Psi^{\beta}(x,z) \ \  .  
\ee
This is a $U(Q)$ gauge field defined on the dual torus (of size
$\frac{1}{l_0} \times \frac{1}{l_1} \times \frac{1}{l_2} \times
\frac{1}{l_3}$). Using Eq.~\ref{tbcfors} one finds that $\hat{A}_{\mu}(z)$
satisfies a relation analogous to Eq.~\ref{tbc} (exchanging the roles of x and z)
in terms of $\Omega'_{\mu}$. 
Now, one can in terms of this field construct the
field strength tensor $\hat{F}_{\mu \nu}$. The Nahm-transformed gauge 
field has the following properties:
\begin{itemize}
\item $\hat{F}_{\mu \nu}$ is again self-dual.
\item The first and second Chern classes and the ranks for the original
and transformed gauge fields are related through:
{\setlength \arraycolsep{2pt}
\bea
rk(\hat{F}) &=& c_2(F)-\frac{1}{2}c_1^2(F) \\
c_1(\hat{F}) &=& -\int_{T^4}(dz_{\mu} \wedge dx_{\mu})^2 \wedge c_1(F) \\
c_2(\hat{F}) &=& rk(F)+\frac{1}{2}c_1^2(F) \quad .
\eea
}
Thus, in particular,  one sees that the roles of the rank of the group 
and the topological charge are exchanged by the Nahm transformation when
the first Chern class vanishes.
\item From the previous statement it follows that if we start with $SU(N)$
gauge fields (with no twist $n_{\mu \nu}= 0 \bmod N$), then the Nahm
transform is in  $SU(Q)$.
\item The Nahm transformation  is an involution: if we apply it twice we go back to 
the original gauge field. Thus, it can be considered a duality transformation. 
\item If we start with a family of gauge fields depending on some 
parameters, then the Nahm transformation  generates a new set of self-dual gauge 
fields depending on those parameters. Hence, we have induced a mapping between 
the moduli spaces of the gauge field and its transform. 
This mapping is an isometry with respect to the natural metric of these 
moduli spaces. 
\end{itemize}
For a proof of these properties, see Ref.~\cite{pbpvb}.

Now, let us consider the vicinity of a point $z$ in the dual torus. 
We can make a Taylor expansion of the positive chirality solutions
of the Dirac equation in the vicinity of this point:
\be
\Psi^{\alpha}(x,z+\Delta z)= \Psi^{\alpha}(x,z) + \Delta z_{\mu}
\Psi_{\mu}^{\alpha}(x,z)+ \Delta z_{\mu} \Delta z_{\nu}
\Psi_{\mu \nu}^{\alpha}(x,z)+ \ldots
\ee
By plugging this equation into the Dirac equation for $z+\Delta z$ and
equating powers of $\Delta z_{\mu}$ on both sides, we obtain for the first two orders:
{\setlength\arraycolsep{2pt}
\begin{eqnarray}
\label{normaleq}
\hat{\bar{D}}_z \Psi^{\alpha}(x,z) & = & 0 \\
\label{main}
\hat{\bar{D}}_z \Psi_{\mu}^{\alpha}(x,z) & = & 2 \pi \imath\,
\bar{\Gamma}_{\mu} \Psi^{\alpha}(x,z) 
\end{eqnarray}}
Now, in terms of these functions, and defining:
{\setlength\arraycolsep{2pt}
\bea
P^{\alpha \beta}_{\mu}(z) & \equiv & \langle \Psi^{\alpha} | \Psi_{\mu}^{\beta} \rangle 
\equiv \int d^4x \left(\Psi^{\alpha}(x,z)\right)^+ \Psi_{\mu}^{\beta}(x,z) \nonumber \\
Q^{\alpha \beta}_{\mu \nu}(z) & \equiv & \langle \Psi_{\mu}^{\alpha} | \Psi_{\nu}^{\beta} \rangle 
\label{prodesc}
\eea}
one can write the vector potential and
the field-strength tensor coming out of the Nahm transformation as follows:
{\setlength\arraycolsep{2pt}
\begin{eqnarray}
\hat{A}_{\mu}(z) & = & \imath P_{\mu}(z) \\ 
\hat{F}_{\mu \nu}(z) & = & \imath \left(Q_{\mu \nu}(z)- Q_{\mu \nu}^{+}(z)+\lbrack P_{\mu}(z),P_{\nu}(z) \rbrack \right)
\label{fmunu}
\end{eqnarray}}
The normalization conditions  imply that $\hat{A}_{\mu}(z)$ is hermitian.
Henceforth, to obtain  the Nahm-transformed gauge field at one point $z$ of the
dual torus, one has simply to solve Eqs.~\ref{normaleq}-\ref{main}.
None of the two equations has a unique solution. Choosing a  solution
within both sets of equations amounts  to a choice of gauge for the 
Nahm-transformed fields. More specifically, any normalized set of solutions  of
Eq.~\ref{normaleq} is  related by a unitary transformation
$\Omega'_{\alpha \beta}(z)$ to any other. Furthermore, once this
choice is made, it can be easily shown that selecting a particular solution
$\Psi_{\mu}^{\alpha}(x,z)$ of  Eq.~\ref{main} amounts to the choice of a
gauge in the neighbourhood of $z$. One can obtain any  solution of
Eq.~\ref{main} by adding to $\Psi_{\mu}^{\alpha}(x,z)$ a general solution of
 Eq.~\ref{normaleq}:  $\Psi^{\alpha}_{\mu}(x,z)+ S_{\mu}^{\beta \alpha}\
\Psi^{\beta}(x,z)$, where $S_{\mu}$ must be antihermitian due to the
normalization conditions. This produces a change in the vector potential
$\hat{A}_{\mu} \longrightarrow \hat{A}_{\mu} + \imath \, S_{\mu}$, but as
can be readily verified from the equations above, $\hat{F}_{\mu \nu}$ is left invariant.
It is in principle possible to impose some set of conditions on the
solutions in order to select a particular gauge for $\hat{A}_{\mu}$.
However, in this paper we will 
concentrate on gauge invariant quantities and, hence, any solution will do. 

In the following sections we will describe how we have been able to 
numerically construct the Nahm transform of a given self-dual gauge field
configuration on the torus by  finding the solutions of
Eqs.~\ref{normaleq}, \ref{main} using the lattice formulation of the theory. 
In the next section we will describe the numerical technique and in the
following we will apply our construction to some explicit examples.

\section{Solving the homogeneous and inhomogenous Dirac equation on the
lattice}

In this section we will explain how we can actually obtain the solutions of
Eqs.~\ref{normaleq}-\ref{main} by studying the problem on the lattice.

All studies
of the Dirac equation on the lattice have to deal with the well-known problem
of fermion doubling. This usually means that  one must either break chiral invariance
explicitly by using Wilson fermions or maintain it at the  expense of
producing spureous lattice solutions to the equation. A middle way is
represented by the use of staggered fermions. 
To put Dirac equations on the lattice we have used both the Wilson action, keeping the value of 
the Wilson parameter $r$ free, 
and a naive fermion action; in both cases, we set the lattice fermion mass equal to zero. 
The second case allows for Weyl (2-component) spinors, while the first one requires
Dirac (4-component) spinors. The results we report here have been obtained with Wilson fermions, so we restrict
our discussion to this latter case.

Our Wilson-Dirac operator reads (the superscript $L$ stands for \emph{lattice}):
{\setlength\arraycolsep{1pt}
\bea
\label{latdirop}
\not\!\!D^L \psi(n) & = 4r\psi(n)-
 \frac{1}{2} \sum_{\mu} \lbrack & (r-\gamma_{\mu}) U_{\mu}(n) \psi (n+\hat{\mu}) + 
\nonumber \\
 & & +(r+\gamma_{\mu})  {U_{\mu}}^{+}(n-\hat{\mu}) \psi (n-\hat{\mu}) \rbrack \quad ,
\eea}
where $U_{\mu}$ is the gauge field, $\hat{\mu}$ is a unitary lattice vector in the $\mu$ direction,
and $\gamma_{\mu}$ are Dirac matrices. 
To add a constant $U(1)$ potential, as required by the construction of the Nahm transformation, we exponentiate the
whole $A_{\mu}+2\pi z_{\mu}I$ to build link variables. Thus, ${\not\!\!D}^L_z$ has the same form as
above but with $U_{\mu}$ replaced by $\exp (-2\pi \imath a z_{\mu}) U_{\mu}$, where $a$ is the lattice spacing.
Note that the argument of the exponential is conveniently dimensionless, as the coordinates $z_{\mu}$ in
the dual space have dimensions of inverse length.

Our first goal
is to solve numerically homogeneous Dirac equations in a given background lattice gauge configuration.
For this purpose, we construct the hermitian positive operator $(\gamma_5 \not\!\!D^L)^2$, for which
we expect to identify a lowest eigenspace of smooth modes, whose degeneracy
must be equal to the one given by the index theorem. The elements of this space can be identified as
a lattice approximation to the continuum zero modes.
Although in general no exact fermionic zero modes will live in the lattice, 
the lowest eigenvalue 
of $(\gamma_5 \not\!\!D^L)^2$ must approach zero with finer discretization
to signal the recovery of continuum behaviour.
All this requirements are indeed verified by our results, and thus we translate the problem of solving
homogeneous Dirac equations into that of finding out these lowest eigenvectors.

Other problem we face when computing solutions is that of keeping the right chirality for the 
fermion fields. Our technique to achieve this proceeds by tuning the value of $r$. 
There is a convenient window within the interval $\lbrack 0,1 \rbrack$ in which $r$ is large enough 
to push up the doublers, 
and small enough so
as to maintain approximate chirality for the resulting solutions. This range of values can be determined by
computing with different $r$, then checking the properties of the result with different 
criteria~\cite{Proc97,FermionZM}. 
A typical value in our computations is $r=10^{-2}$.

We are finally able to obtain vectors that are approximately chiral, the ``wrong'' half (chirality
opposite to the one dictated by the sign of the topological charge of the gauge configuration) of
the spinor being numerically negligible everywhere with respect to the ``right'' half. Above the lowest space, 
whose degeneracy is given by the index theorem, the typical structure for
the lowest part of our spectra presents a first excited continuum level, the order of magnitude of its eigenvalue
being stable with changing lattice volume, and a set of doubler states between this excited level and the zero modes level. 
We can recognize the doublers by the oscillating behaviour
of their invariant densities, as well as by comparison with the known doubling symmetry pattern. Moreover,
the evolution with $r$ of the respective eigenvalues for the quasi-zero modes and the doublers
is qualitatively different.
The gap between the approximate zero modes level and the doubler levels is determined essentially by the value
of $r$.

To compute lowest eigenvectors of the operator $(\gamma_5 \not\!\!D^L)^2$
we have used a standard conjugate gradient (CG) method, basically the one described
in~\cite{KalkCG} (without implementing the acceleration method described in this reference), which gives
eigenvectors of the hermitian, positive definite operator $A$ with increasing eigenvalue by minimizing the Ritz functional:
\be
\mu(v)=\frac{<v,Av>}{<v,v>} \quad ,
\ee
with $v$ a trial CG vector.

To check by consistency our CG method we have performed alternative calculations using both a Lanczos procedure and
a minimization technique inspired by the cooling algorithm used in lattice Yang-Mills studies. The results for the
zero modes in a number of qualitatively different gauge configurations will be the subject of future 
publication~\cite{FermionZM}.

To solve inhomogenous Dirac equations of the form ${\not\!\!D \psi}=\phi$, as required by our construction 
of the Nahm transformation 
through a local Taylor expansion, we use a variant of the CG method, 
the stabilized biconjugate gradient algorithm (BiCGStab)~\cite{BiCGorig}, whose adaptation for the purpose of
inverting lattice Dirac operators is described in~\cite{BiCG}. We have
used directly the implementation described in this latter reference, with minor operative changes. The basic idea of the
construction is the following: if we apply CG to an inhomogenous equation of the form $Av=w$, for each iteration we
will have a trial solution $v_i$, a search direction $p_i$ and a residue $r_i=w-Av_i$; the new trial vector is
constructed as $v_{i+1}=v_i+\alpha p_i$, with $\alpha$ determined by the requirement that the new residue has
minimum norm, and the problem is to find the new $p_{i+1}$ (which fulfills the same condition for search directions
as in ``raw'' CG) and $r_{i+1}$ with the lowest computational cost, 
which essentially
means with a lowest number of operator-vector multiplications. The BiCG method incorporates an additional sequence
of vectors $\hat{r}_i$ such that $r_j^{+} \hat{r}_i = \hat{r}_j^{+} r_i=0$ for
$i=0,1,\ldots,j-1$. With this construction, the residuals satisfy
\bea
r_j=P_j(A)r_0 & , & \hat{r}_j=P_j(A^{+})\hat{r}_0 \quad ,
\nonumber
\eea
where $P_j$ are polynomials of degree $\le j$. If the method converges, these polynomials can be identified as
`reductors' for the norm of the residue.

This convergence behaviour can be enhanced as in the conjugate gradient squared (CGS) algorithm, which furthermore 
avoids the computation of the $\hat{r}_i$ and makes residuals satisfy $r_j=P_j^2(A)r_0$, with the same $P_j$. 
However, the convergence of CGS is usually far from being smooth, and the possibility exists of obtaining large
peaks for $|r_i|$ which can plague a systematic application of the method. The BiCGStab cures this problem by
making $r_j=Q_j(A)P_j(A)r_0$, with an appropriate form for the polynomial $Q_j$ so as to keep
a good reduction behaviour while keeping in a smoother regime the evolution of $|r_i|$.
Further technical details can be found in the references.

The application
of this method to our problem allows us to efficiently compute approximate solutions, with typical values for 
$|\phi-{\not\!\!D \psi}|$
far below the corrections to continuum quantities eventually coming from other sources, thus not adding 
significative new errors to the calculations in which these vectors enter.

\section{Application to the case of non-trivial $SU(2)$ gauge field
configurations on the torus}
In order to test the applicability and stability of our method we have
first studied a case for which an explicit analytical solution is known. 
This occurs for gauge configurations with constant field strength. The Dirac 
equation in this background can be solved by techniques which are very
similar to the ones used to   study small deformations around these 
gauge fields \cite{vanbaal6}. An explicit example has been worked out in
Ref.~\cite{pvb5}. For comparison we take precisely this case in our numerical 
work. 

Our gauge configuration has twist tensor defined by $n_{03}=n_{21}=2$, the other
components being zero; its topological charge is 2. The gauge potential and the field strength 
tensor can be written as:
\bea
A_{\mu} &=& -\frac{1}{2}\pi \frac{n_{\mu \nu}}{l_{\mu}l_{\nu}} x_{\nu} \sigma_3 \\
F_{\mu \nu} &=& \pi \frac{n_{\mu \nu}}{l_{\mu}l_{\nu}} \sigma_3 \quad ,
\eea
where $l_{\mu}$ are, as defined before, the lengths of the torus,
which we set equal to 1. (Note that this $F$ is actually anti-self-dual).
The Nahm transformation can be carried out analytically in this case~\cite{pvb5},
and the result is $\hat{A}_{\mu}=-A_{\mu}$, $\hat{F}_{\mu \nu}=\tilde{F}_{\mu \nu}=-F_{\mu \nu}$ with the
natural gauge choice for the transformed fields.

To compare with this result, we first put the configuration on the lattice, which can be
done exactly by exponentiating the above form for $A_{\mu}$ to build links. Now, the
required numerical solutions to lattice Dirac equations can be computed, and the transformed
field extracted from their scalar products according to  our procedure. To be able to obtain
a large set of transformed points,
we have worked in a $8^4$ lattice. At the end we get a perfect structural reproduction of 
the exact results, with only a slight error for the field value. The only nonvanishing
fields for our numerical transformation are $E_3$ and $B_3$, as required, and self-duality
is verified almost exactly within the machine numerical precision; when a gauge is
fixed (by directly diagonalizing the fields at each {\em z}--point, because the gauge fixing procedure
we describe below for nontrivial configurations does not make sense in this case), we end up with:
\be
E_3=B_3=1.05 E_3^C \quad ,
\ee
with $E_3^C$ the continuum field. The 5 \% difference
must come from the discretization corrections to the solutions
of Dirac equations, which are sizable due to the small number of lattice points.
In any case, we can conclude that our numerical implementation of the Nahm 
transformation is working satisfactorily for this test example.

\vspace{1cm}

Now let us consider another case for which no analytical solution is known. 
Motivated by our own interest in this type of configurations, our starting
point is a self-dual $SU(2)$ gauge field configuration with non-trivial twist 
and fractional topological charge. In particular, we take a solution having 
non-orthogonal twist given by the twist tensor:
\be
n_{0 i } = ( 1, 0 ,0)\ \ \  \frac{1}{2} \epsilon_{i j k}\, n_{j k}= ( 1, 1,  1 )
\ee
Given this twist, the topological charge must be a half integer value, equal
to the action divided by $ 8 \pi^2$. We are interested in the absolute
minimum action solution in this twist sector. This has action $4 \pi^2$ and
topological charge $\frac{1}{2}$. A configuration of this type, which is
unique modulo space-time  translations and discrete symmetries, is easy to
obtain numerically by minimizing the lattice action in this twist sector
(see, for example, Refs.~\cite{gpg-as,gpg-a}). The action density consists on
a single lump with size of the order of the smallest torus length.
For reasons that will be clear soon we took the torus sizes in the
ratio $l_0:l_1:l_2:l_3\ = 2:1:2:1$.

\begin{figure}[!h]

\caption{Action density for the original lattice gauge configuration in the plane of the four absolute maxima.
The total action has been normalized to~1.}

\vbox{ \hbox{ \hskip40pt \epsfxsize=300pt \hbox{\epsffile{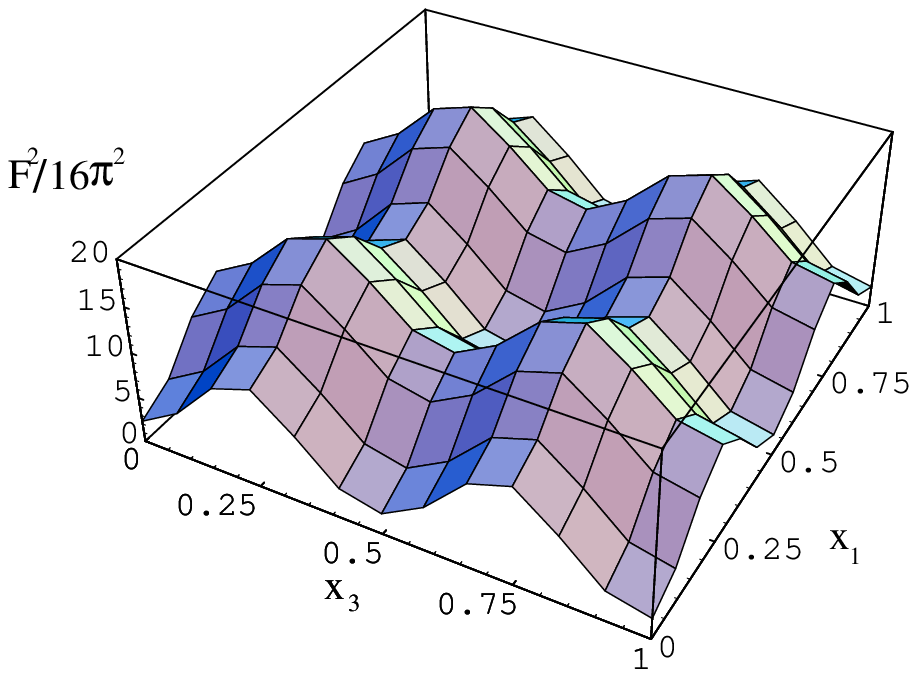} } }   
     }

\label{OrigConf}

\end{figure}

However, $SU(N)$ configurations with twist
are not good starting points for the Nahm transformation since these boundary
conditions are singular  for  spinors transforming in the fundamental
representation. There is a number of ways out of this problem. We will
choose the following one. Instead of considering a single torus period we
will consider more than one period in the short directions. The resulting
configuration, after applying an appropriate gauge transformation, has no twist
or singularity in the new torus of size $l_{\mu}=1$. Gauge invariant
quantities are periodic with period $\frac{l_{\mu}}{2}$ in the $1$ and $3$
directions. Hence the action density consists on four lumps like those
described previously and the total topological charge and action are $2$ and
$16 \pi^2$ respectively. In Fig.~\ref{OrigConf} we plot the action density in the
$x_1-x_3$ plane setting the other two coordinates to the value giving
maximal action density. This is the original gauge field
configuration to which we will apply the Nahm transformation. This will again be a
self-dual $SU(2)$ configuration with topological charge equal to $2$.

In the numerical construction of the $Q=1/2$ lump we have worked in a lattice of size
$12 \times 6 \times 12 \times 6$. After ``glueing'' the lump to itself to obtain
the configuration explained above, the result
is a $12^4$ lattice, where all the subsequent computations are carried out.

\begin{figure}[!h]

\caption{Invariant densities for the two orthogonal zero modes chosen for the original lattice configuration,
in the plane of the four absolute maxima. Both have norm 1 in the continuum.}

\vbox{ \hbox{ \hskip70pt \epsfxsize=250pt \hbox{\epsffile{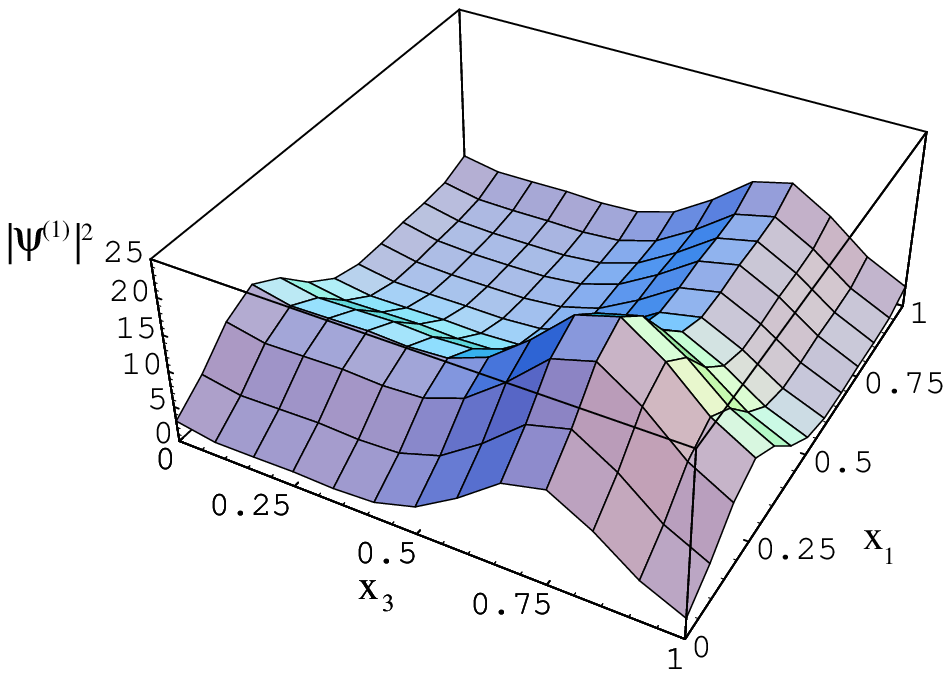} } }   
     }

\vbox{ \hbox{ \hskip70pt \epsfxsize=250pt \hbox{\epsffile{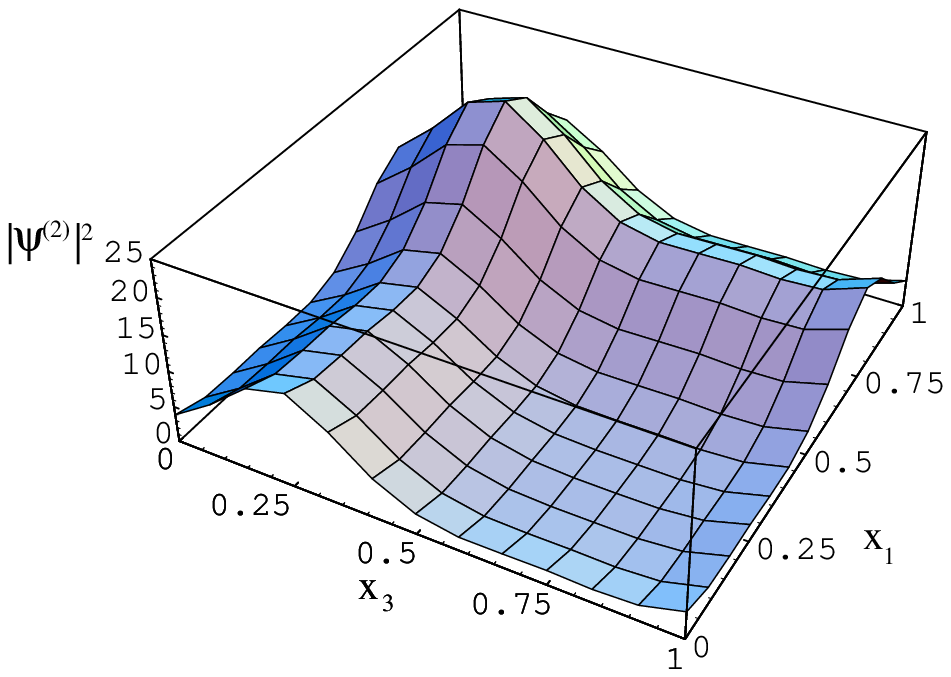} } }   
     }

\label{ZMs}

\end{figure}

The first thing we did was to look at the solutions of the Dirac equation
in the background of this configuration. Indeed on the lattice there are no
exact solutions. What we actually did, as explained in the previous chapter,
was to find the smallest eigenvalue of the operator $(\gamma_5 \not\!\!D^L)^2$ and the
corresponding eigenvectors.   There are 2 linearly independent degenerate 
solutions. We will choose the following procedure to choose a basis in this
two-dimensional space (which fixes the gauge for $\hat{F}$). Consider one of the points $x_0$ on the lattice
where the action density peaks (there are 4 points of this type). Now we
choose the basis of solutions of the Dirac equation such that the hermitian
matrix $M_{\alpha \beta}(x_0)$ is diagonal:
\be
M_{\alpha \beta}(x_0) \equiv \left(\Psi^{ \alpha}(x_0)\right)^+ \Psi^{\beta}(x_0) =
\left(\begin{array}{cc} \lambda_1 & 0\\ 0 & \lambda_2\end{array} \right)  \ \ .
\ee
It is a scalar product (sum over spin and color indices) of the $2N$-dimensional vectors $ \Psi^{
\alpha}(x_0)$. We choose $\lambda_1 > \lambda_2$. This condition does not
fix completely the solutions, one is still free to multiply each element of
the basis by an independent  phase. We fix the remaining arbitrarity
by imposing $M_{\alpha \beta}(x_0')$ to be real, where $x_0'=x_0+\frac{1}{2}\hat{l}_1$ is another of the
points where the action density peaks. There is an overall choice of phase
which will not affect our results. Our first result will be to give the
density matrix $M_{\alpha \beta}(x)$, which is a (gauge invariant with respect to transformations
of the original gauge field) $2 \times 2$ hermitian
matrix  field defined on the original torus. In Fig.~\ref{ZMs} we show the value of
the corresponding densities $M_{1 1}$ and $M_{2 2 }$ in the same  plane
where  we gave the action density of the configuration before. Notice that
the first solution peaks precisely at a point $x_0$ where the action
density peaks, and was used to fix a basis in the space of solutions.
On the contrary the density of the other solution  $M_{2 2 }$ shows the same
behaviour but displaced by a vector $(0,\frac{1}{2}, 0, \frac{1}{2})$ in
space. Other interesting property of the solution is that the matrix
$M_{\alpha \beta}(x)$ turns out to be real at all space points.   

The next step was  to study the solutions of the Dirac equation for other
values of $z$.  We use the same conditions to choose a basis in the space of 
solutions. The nice symmetry properties of the $z=0$ solutions are lost for 
arbitrary  values of $z$, but the solution seems to show a nice smooth
behaviour which makes it reliable  to consider them a good approximation to
the continuum limit ones.  

Now, we consider the solutions of Eq.~\ref{main}, where the left hand-side
is fixed by the solutions of Eq.~\ref{normaleq}, as explained before. The
solution is now unique, for a given choice of  the direction ${\mu}$ and of the
index $\alpha$. This is in contrast with the continuum case and is precisely
due to the fact that the lattice operator has no exact zero-modes.
Fortunately, we are interested in $\hat{F}_{\mu \nu}$ which is insensitive to the
continuum solution chosen, hence we feel satisfied with the unique solution
provided by the lattice. For all values of $z$ studied, the resulting
$\Psi_{\mu}(x,z)$ obtained with this procedure look pretty smooth,
suggesting that indeed we are dealing with lattice approximants to the
continuum functions. Finally by use of Eqs.~\ref{prodesc},~\ref{fmunu} we can calculate the
components of $\hat{F}_{\mu \nu}$. 

The first thing to check is that
the result obtained is indeed a self-dual field-strength tensor. For all the $z$
values studied this was very approximately the case. In Table~1 we show
the results for some selected values of $z$. The departures from self-duality were always found
to be at most of order $10^{-2}$,
 as for the original lattice gauge configuration. 
We have actually explored points in
different regions of the dual torus (periodicity in $z_{\mu}$ was checked
and found to work extremely well) and always got results which indicate regularity as
a function of $z$. 

\begin{figure}[!h]

\caption{Action density for the transformed gauge configuration in the $(z_0=1/2,z_2=1/2)$ plane. The normalization
is the same as in the original lattice configuration. The center in this plane has been displaced to the
origin of coordinates for clarity, and
an interpolating procedure has been used to obtain a smooth surface. 
}

\vbox{ \hbox{ \hskip40pt \epsfxsize=300pt \hbox{\epsffile{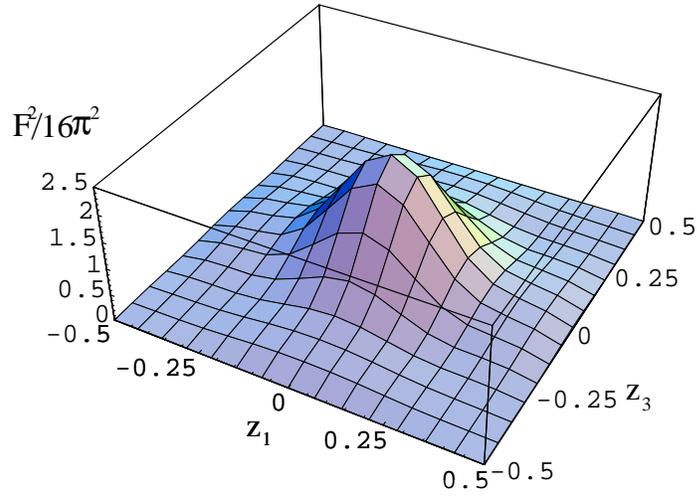} } }   
     }

\label{zPlane}

\end{figure}

In Fig.~\ref{zPlane} we show a plot for the action density obtained in the
plane $(z_2=1/2,z_4=1/2)$, where a single peak appears. A total of 20 points in this plane have been generated and
we used the symmetry under $\frac{\pi}{2}$ rotations around the center 
of this plane to increase the statistics of the surface. This symmetry has
indeed been checked and works to a high level of precision. 

\begin{figure}[!h]

\caption{Comparison of action densities for the Nahm transformed field (dots) and the original lattice field (solid line)
along a line with two absolute maxima. The line has been extracted from a fit to the lattice points densities. The
center has been displaced to the origin of coordinates for clarity.}

\vbox{ \hbox{ \hskip75pt \epsfxsize=300pt \hbox{\epsffile{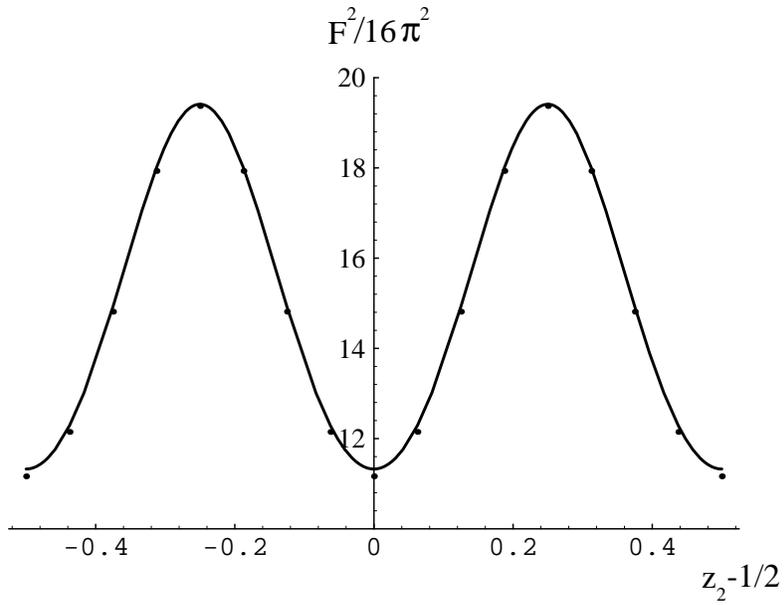} } }   
     }

\label{Compara}

\end{figure}

The action density of the Nahm-transformed solution resembles very much the
action density of the original configuration. When comparing one has to take
into account that the centers (the actual maxima of the action density) of the original
configuration do not lie
exactly at a lattice point. To make a comparison which would be more
precise, we made a fit to the action density of the original lattice
configuration using a few Fourier components in each variable and this
allowed us to interpolate the information of the original lattice
configuration to points lying in between the lattice points. In Fig.~\ref{Compara} we
compare the result of this fit (solid curve) to the actual data points
obtained for the transformed field along one of the straight lines in the torus
joining two centers. We emphasize that the curve is obtained by fitting  the original
configuration and not the Nahm-transformed one. In all the points explored we
always found that the action density of the transformed field matches quite
precisely the one of the original configuration. We proceeded to investigate
systematically what is the exact relation between the original configuration
and its  Nahm transform by looking at the rest of the gauge invariant local
quantities. These are the traces of the products of two spatial components
of the electric field $Tr(E_i(z)\,E_j(z))$. We made a careful selection
of the $z$ points to distinguish the different possibilities.
The differences between the original and transformed action densities in this
comparison can grow up to 
5-10 \%, which can be partly due to small uncertainties in the determination of
the peak centers for the lattice configurations, from which the direct coordinate
comparison depends.
However, we found a consistent way to recognize directions by
looking at the signs of the traces of crossed products of fields.
Finally, we arrived at the conclusion that, up to a gauge, the relation between 
the original and transformed field tensors is as follows:
{\setlength \arraycolsep{2pt}
\bea
\hat{F}(z) &=& F(x'(z))\\
x'(z_0,z_1,z_2,z_3) &=& (z_1,-z_0,z_3,-z_2) 
\eea
}
Actually, changing $z \rightarrow -z$ is also possible since the configuration 
is invariant under this transformation.
To show an explicit example of the 
comparison we applied the Nahm transformation precisely at a transformed 
lattice point. The result is shown in Table~2, where the traces
$Tr(E_i(x)\,E_j(x))$ are compared for two corresponding $x$ and $z$ points.
To arrive to similar results for  $A_{\mu}$, we should be able to compute
and compare  Polyakov lines. This is computationally very expensive with the present
method, although some variation of it could be particularly helpful in this respect
(cf. comments at the end of the concluding section).

\vspace{1cm}

To finish this section we want to mention another application of our numerical methods to the
study of a given gauge configuration.
It comes from the study of fermion zero modes in the adjoint representation of the
gauge group, which can be directly adressed with our techniques. In this case there is
no need to have strictly periodic gauge configurations, and we can consider also twisted
background fields. The index theorem indicates that $2NQ$ zero modes are to be found
in this case. Furthermore, in contrast to the case of the fundamental
representation, a new important property is expected: the supersymmetry present
in the theory with adjoint fermions supplies 2 of the zero modes as supersymmetric transforms
of the background (anti)self-dual gauge field~\cite{CohenGomez}, through
\be
\psi^{A a}_{\alpha}(x)=\frac{1}{2}\sigma_{\mu \nu}^{AB}F_{\mu \nu}^a(x) u_{\alpha}^B \quad ,
\label{SUSY}
\ee
where 
$a$ is an adjoint color index in the Pauli basis, $\alpha$ an index in the space of zero modes, $A,B$ are
spin indices, $u$ is the spinor
parameter of the transformation, and 
$\sigma_{\mu \nu} \equiv \frac{\imath}{4}(\Gamma_{\mu}\bar{\Gamma}_{\nu}-\Gamma_{\nu}\bar{\Gamma}_{\mu})$.

Our lattice computations are again able to reproduce these continuum results~\cite{Proc97,FermionZM}. 
In this case, the expected $2NQ$-dimensional
space of lowest modes actually splits into $NQ$ spaces of dimension 2, above which doubler levels appear.
Each couple of zero modes present a geometrically distinct character, and the space related to the
`supersymmetric' zero modes can be easily identified, typically being the highest one. This allows
to extract numerical approximations for the \emph{continuum} field $F_{\mu \nu}$, which are expected to
differ from the lattice values computed from plaquettes, 
as finite lattice spacing corrections enter these two
quantities in completely different ways (a complete analysis of discretization errors in this new
approach is still lacking). We have verified that this is indeed
the case, the difference being most appreciable around absolute maxima and minima for the action
density. However, the results for the gauge field coming from adjoint zero modes still compares successfully
(although with larger errors) with the Nahm-transformed field when checking the correspondence between them.
Further study is required in order to understand the differences in lattice corrections and to obtain a better
way to compare the results consistently.

\section{Conclusions and future outlook}
In this paper we have presented a way in which one can numerically
construct the Nahm transform of a self-dual Yang-Mills classical 
configuration discretized on the lattice. This opens the door for a 
more thorough study which will allow to disentangle the main features of this
transformation. For example, one can study  fixed points of the transformation,
and relate solutions belonging to different gauge groups. This goal  however 
requires an important computational effort and at the moment we are
improving our algorithm and codes to make this a feasible task. An analysis of systematic
errors is also underway. The analytic 
information on the transform can be of great help in suggesting the
appropriate choice of initial configurations and points to study.

We have applied the Nahm transformation to a self-dual gauge field
configuration built out of one carrying non-orthogonal twist. We found 
that in this case the transformation gives back the original configuration 
at a transformed point in space-time. This sort of constraint could turn 
out to be useful in finding  the analytical solution.

To conclude,  we want to  comment on a point which was left out
and will be  developed in future publications. This is the fact that one can
use techniques similar to the ones described to obtain a lattice Nahm
transformation (that is, a lattice gauge configuration which is the Nahm transform
of another lattice gauge configuration) by directly constructing link variables instead
of continuum ones. This has a number of advantages as a parallel method; for example,
Polyakov loops for the Nahm-transformed configuration can now be easily computed 
from lattice links.

The construction proceeds as follows. Let us
first select a grid on the dual torus $z_{\mu}=\tilde{a} n_{\mu}$, where $\tilde{a}$
is the spacing in this new lattice. The solutions $\Psi^{\alpha}(x,z)$ of Eq.~\ref{normaleq}
can now be computed in exactly the same way we have described. Once they are known, form
the matrices:
\be
{\cal{V}}_{\mu}^{\alpha \beta}(z)=\sum_{x} \Psi^{\alpha +}(x,z) \Psi^{\beta}(x,z+\hat{\tilde{\mu}})
\ee
where color and spin indices are summed over and $\hat{\tilde{\mu}}$ is a unitary vector in the $z$-lattice.
These matrices transform as lattice links under a $z$-dependent change of basis in the space of zero modes, 
i.e. a gauge transformation in $z$-space:
{\setlength \arraycolsep{2pt}
\bea
\Psi^{\alpha}(x,z) & {\longrightarrow} & \Psi'^{\alpha}(x,z) = \Psi^{\beta}(x,z) \Omega^{+ \beta \alpha}(z) \\
{\cal{V}}^{\alpha \beta}_{\mu} & {\longrightarrow} & {\cal{V}}'^{\alpha \beta}_{\mu}=\Omega^{\alpha \gamma}(z) {\cal{V}}^{\gamma \delta}_{\mu} \Omega^{+ \delta \beta}(z+\hat{\tilde{\mu}})
\eea
}
but are not unitary, as required to be identified as link variables. To extract a unitary matrix from $\cal{V}_{\mu}$
while keeping the transformation property we can decompose it as the product of a unitary matrix
times a hermitian matrix~\footnote{For $Q=2$ there is an easier way: the matrix 
${\cal{U}}_{\mu} \equiv {\cal{V}}_{\mu} + \sigma_2 {\cal{V}}_{\mu}^{*} \sigma_2$ has the
property ${\cal{U}}_{\mu} {\cal{U}}_{\mu}^{+} = det({\cal{U}}_{\mu})I$, and it is straightforward
to extract a $SU(Q)$ matrix from it.}~, ${\cal{V}}_{\mu}(z)=H_{\mu}(z) U_{\mu}(z)$. Finally, a
$SU(Q)$ matrix has to be extracted from each $U_{\mu}$, which in general is in $U(Q)$.
There is a number of ambiguities here, having to do with remaining freedoms in the `unitarization'
process, but these can in principle be dealt with by appealing to continuity arguments when going
from one $z$ point to the following, once a convenient
gauge has been chosen (for which the method we have described in the preceding section remains valid).

It can be seen that the naive continuum
limit for the new link variables is indeed the continuum field in Eq.~\ref{NT}, although a rather complicated
behaviour arises for lattice corrections, which makes it difficult to keep them controlled.
This latter point seems to be the main difficulty to exploit this method. Further study, and probably
use of larger lattices is required before it can be efficiently implemented.

\section*{Acknowledgements}
This work was financed by the CICYT under grant
AEN97-1678. An important part of the numerical calculations
which are here presented have been carried on the Alphaserver8400
at the Cen\-tro de Com\-pu\-ta\-ci\'on Cien\-t\'{\i}\-fi\-ca (UAM).
We thank \'Alvaro Montero for help in the construction of gauge configurations
and in the development of numerical methods.

\newpage

\newpage

\newpage

\newpage

\newpage

\begin{table}[!h]

\label{tb:TabSd}

\begin{center}
\caption{Values for the transformed electric and magnetic gauge fields in some selected points of the
$z$-space are given. The number of significative figures is dictated by the deviations from self-duality
appearing in the original lattice configuration. The first point corresponds to an absolute maximum;
the second, to the maximum in a plane in which $E_1=E_3=0$; the third, to a point with no special
character.}

\vspace{1cm}

\begin{tabular}{|c|c|c|}
\hline 
{\em z}--space point &  $E_1^a,E_2^a,E_3^a$ & $B_1^a,B_2^a,B_3^a$ \\   \hline \hline
 
$(1/2,1/4,1/2,1/4)$ & 
$(-0.049,0.012,-1.262)$ & 
$(-0.049,0.012,-1.262)$ \\
 &
$(-1.204,0.444,0.051)$ & 
$(-1.212,0.447,0.051)$  \\
 &
$(-0.436,-1.182,0.005)$ & 
$(-0.436,-1,182,0.005)$ \\

\hline

$(1/2,1/2,1/2,1/2)$ & 
$(0.000,0.000,0.000)$ & 
$(0.000,0.000,0.000)$  \\
 &
$(0.218,-0.705,0.000)$ & 
$(0.225,-0.725,0.000)$  \\
 &
$(0.000,0.000,0.000)$ & 
$(0.000,0.000,0.000)$  \\

\hline


$(3/4,1/2,1/2,2/3)$ & 
$(0.031,-0.001,-0.262)$ & 
$(0.032,-0.002,-0.262)$  \\
 &
$(-0.311,-0.166,-0.038)$ & 
$(-0.316,-0.169,-0.037)$  \\
 &
$(0.076,-0.287,0.010)$ & 
$(0.076,-0.287,0.010)$  \\

\hline

\end{tabular}

\end{center}

\end{table}

\begin{table}[!h]

\label{tb:TabCorr}

\begin{center}
\caption{Values for the traces of products of electric field components for corresponding $x_0$ and $z_0$ points:
$x_0=(0.312,0.584,0.596,0.766)$, $z_0=(0.584,0.688,0.766,0.404)$.
$x_0$ corresponds to a lattice point, selected such that clear hierarchies among the invariants are established.}

\vspace{1cm}

\begin{tabular}{|c|c|c|}
\hline 
 &  $x_0$ & $z_0$ \\   \hline \hline
$Tr(E_1(x)\,E_1(x))$ & 0.418  & 0.460  \\
$Tr(E_2(x)\,E_2(x))$ & 0.465  & 0.488  \\
$Tr(E_3(x)\,E_3(x))$ & 0.383  & 0.412  \\
$Tr(E_1(x)\,E_2(x))$ & 0.033  & 0.031  \\
$Tr(E_1(x)\,E_3(x))$ & -0.027  & -0.027  \\
$Tr(E_2(x)\,E_3(x))$ & 0.024  &  0.025 \\

\hline

\end{tabular}

\end{center}

\end{table}

\end{document}